\documentclass[11pt,twoside]{article}

\usepackage{asp2006}

\begin{document}
\title{GHaFaS: Galaxy Halpha Fabry-Perot Spectrometer\\
for the WHT}   
\author{C. Carignan and O. Hernandez}   
\affil{Laboratoire d'Astrophysique Exp\'erimentale, D\'epartement de physique,
Universit\'e de Montr\'eal, Montr\'eal, Qu\'ebec, Canada}    
\author{J. E. Beckman and K. Fathi}
\affil{Instituto de Astrof\'\i sica de Canarias, E-38205 La Laguna, Tenerife,
Spain}

\begin{abstract}
A new scanning Fabry-Perot system will soon be available at the Nasmyth focus
of the 4,2m William Hershell Telescope (WHT). It has been designed by the 
Laboratoire d'Astrophysique Exp\'erimentale (LAE) in Montr\'eal and is being
built in collaboration with astronomers at the Instituto de Astrof\'\i sica de 
Canarias (IAC). The instrument will see first light at the beginning of July 2007.
\end{abstract}

\vspace{-0.5cm}
\section*{Description of the Projet}

GHaFaS is an improved version of the scanning Fabry-Perot system FaNTOmM
\citep{he2003}, which is a resident instrument on the
Observatoire du mont M\'egantic (OMM) 1,6m telescope and which has also
been used on the CFH and ESO La Silla 3,6m telescopes. The complete 
system is composed of a focal reducer, a calibration unit, a filter wheel
for the order sorter filters, an FP etalon and an IPCS camera. The IPCS
is composed of an Hamamatsu intensifier MCP tube which intensifies
every generated electron coming from the photocatode by a factor 10$^7$.
Each photon event, recorded on a DALSA CCD, is then analysed by a centering 
algorithm. With this amplification, the camera has essentially no readout noise.
Because of this, a zero noise IPCS is to be prefered to CCDs at very low 
flux level \citep{ga2002}, even if the GaAs IPCS has only a DQE of 25\%.
Moreover, because of the fast scanning capability, it can average out the
variation of atmospheric transmission which is not possible with the long
integration times needed per channel for the CCDs in order to beat the
read-out noise.

In the last 3 years with the FaNTOmM system, around 150 galaxies were
observed on the OMM, CFH and ESO La Silla telescopes in the context of
3 large surveys: the SINGS sample \citep{da2006}, a
survey of barred galaxies, the BHabar sample \citep{he2005} and
a sample of Virgo spirals \citep{ch2006}. While the first scientific
justification of the Montr\'eal group was to derive high spatial resolution
optical rotation curves for mass modeling purposes, the data was also used 
by IAC astronomers to constrain the role of gravitational perturbations
as well as feedback from individual HII regions on the evolution of structures
in galaxies and by a Berkeley-Munich group and the G\'EPI group in Paris to
compare those local samples to high z galaxies.

GHaFaS will come with its own custom designed focal reducer developed to be
optically and mechanically compatible with the Nasmyth focus of the WHT.
The system has its own control and data acquisition system. It will have 
a 4x4 arcmin field with a 0.45 arcsec pixel and $\sim$5 km/s velocity
resolution. Full acquisition and reduction software (mainly based on IDL
routines) will be provided by the Montr\'eal group.
The project will be done in 3 phases. For Phase I (July 2007), the optical
system (focal reducer, filter wheel \& calibration unit) will be delivered 
to the WHT and used with the camera FaNTOmM for this first run. For Phase II
(end of 2007), an improved GaAs IPCS will be added to the system. Phase III 
(early 2008) will provide an FP controller and possibly a monochromator to
calibrate the data at the observing wavelength.

\section*{Science to be done with GHaFaS}

Two-dimensional kinematics is a very powerful technique for studying the 
structure and evolution of galaxies. The distribution of dark matter,
circumnuclear star formation and fuelling of active galactic nuclei, detection
of counter-rotating and kinematically decoupled components, and the effects
of interaction between massive stars and the interstellar medium are
among the physical phenomena which can be studied with this technique: see e.g.
\citet{fa2007} and \citet{re2007}.

The first large program for which we plan to use GHaFaS consists at observing 
a sample of 46 carefully selected nearby galaxies which are all included in the
SINGS, THINGS, GALEX, and other CO and optical archives. Due to the angular
size of some of the objects, 2-4 fields may be necessary to reach the 25th 
magnitude radius. This totals to 72 fields which will require $\sim$18 clear
nights of observing with GHaFaS on the WHT. Priority will be given to enlarge our Virgo
sample of galaxies. The full sample ranges from elliptical (with emission)
to irregular galaxies, 2/3 of which are intermediate-type objects, since
this is where highly star-forming regions will be observed. It will be
possible to use this sample for many scientific projects ranging from the 
large scale mass modeling using the velocity fields in order to derive the dark 
matter density profiles to the study of the internal kinematics of the
individual HII regions.


\end{document}